\documentclass[11pt]{article}
\usepackage{fa2025}
\usepackage{amsmath}
\usepackage{cite}
\usepackage{url}
\usepackage{graphicx}
\usepackage{color}
\usepackage[utf8]{inputenc}
\graphicspath{{./fig/}}
\title{Influence of the basins of attraction in the register jumps of the clarinet}

\multauthor
{Nathan Szwarcberg$^{1,2*}$ \hspace{1cm} Tom Colinot$^1$ \hspace{1cm} Christophe Vergez$^2$} { \bfseries{Michaël Jousserand$^1$}\\
  $^1$ Buffet Crampon, 5 Rue Maurice Berteaux, 78711 Mantes-la-Ville, France\\
$^2$ Aix Marseille Univ, CNRS, Centrale Med, LMA, Marseille, France
\correspondingauthor{nathan.szwarcberg@buffetcrampon.com}{Szwarcberg et al.}
}

\begin{document}
\maketitle
\begin{abstract}
When playing the clarinet, opening the register hole allows for a transition from the first to the second register, producing a twelfth interval. On an artificial mouth, the blowing pressure range where the second register remains stable can be determined by gradually varying the blowing pressure while keeping the register hole open. However, when the register hole is opened while the instrument is already producing the first register, the range of blowing pressures that lead to a stable second register is narrower than the full stability zone of the second register.

This phenomenon is investigated numerically by performing multiple hole openings at different times for each blowing pressure value. The evolution of the probability of reaching the second register is computed, and its relationship with the structure of the basin of attraction of the second register is analyzed.
\end{abstract}

\section{Introduction}\label{sec:introduction}

When characterizing a clarinet fingering, one of the first steps consists in measuring the minimum and maximum blowing pressure that play a note, for a fixed embouchure.
The artificial mouth \cite{backus1961vibrations, mcginnis1943experimental, li2016effect, chatziioannou2019investigating} is commonly used to determine these limits, known as the oscillation and extinction thresholds \cite{dalmont2005analytical, dalmont_oscillation_2007, atig2004saturation}, by gradually increasing the blowing pressure.
Above the extinction threshold, the reed is pressed against the mouthpiece and stops vibrating.
When the pressure is then reduced, the reed starts oscillating again at a lower pressure, sometimes called the ``inverse threshold'' \cite{dalmont_oscillation_2007}. 
This difference between the extinction and inverse thresholds creates a hysteresis cycle, where the equilibrium (no sound) and the oscillating regime are multistable \cite{colinot2021multistability,colinot2025cartography}.

The basin of attraction of a regime defines the set of initial conditions that lead to it. 
In a multistable system, knowing these basins helps predict which regime a musician is most likely to play \cite{benade1996physics}. 
However, calculating the full basin of attraction is highly time-consuming due to the high dimension of the phase space. 
Additionally, it is unclear whether a chosen initial condition accurately represents a musician's playing.

This study addresses these challenges by focusing on transitions between two notes. 
In this case, all initial conditions lie on the limit cycle of the regime of the first note.

A well-known transition on the clarinet happens when pressing the register key, which shifts from the first register to the second by an ascending interval of a twelfth.
For beginner clarinetists, practicing this transition is important to avoid unintended notes when opening the hole.

This paper investigates a physical model of a cylindrical clarinet with a register hole. 
To allow the model to reproduce the register transition, nonlinear losses in the register hole are included \cite{szwarcberg2024second, szwarcberg2024oscillation}. 
The model is introduced in Section~\ref{sec:model}.
Time-domain simulations, similar to artificial mouth experiments, are then conducted. 
First, blowing pressure ramps are tested with the register hole both closed and open (Section~\ref{sec:ramps}) to identify stable and multistable regions. 
Next, for blowing pressure values in the multistability range between the second register and the equilibrium, several hole openings are performed to observe how the first register’s limit cycle interacts with the attraction basins of the two competing regimes (Section~\ref{sec:holeopenings}). 
Finally, basin stability is investigated by introducing random perturbations when opening the register hole \cite{menck2013basin}.

\section{Numerical model}\label{sec:model}
\subsection{Digital resonators}\label{sec:model resonator}

\begin{figure}
	\centering
	\includegraphics[width=.5\textwidth]{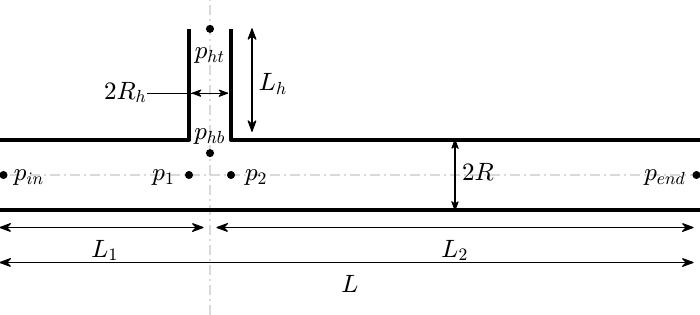}
	\caption{Definition of the digital resonators studied.}
	\label{fig:2}
\end{figure}

The digital resonator is presented on Figure \ref{fig:2}.
It is composed of a first tube of length $L_1=132$~mm, radius $R=7.5$~mm and cross-section $S=\pi R^2$.
The characteristic impedance of plane waves  propagating through the tube is $Z_c=\rho_0 c_0/S$ where $\rho_0=1.23~\mathrm{kg\cdot m^{-3}}$ and $c_0=343~\mathrm{m\cdot s^{-1}}$.
The acoustic field in the first tube is described by the pressure at the left extremity $p_{in}$, and at the right extremity $p_1$.

The tube is branched to a side hole of length $L_h$~$=$~$12.7$~mm, radius $R_h=1.5$~mm, cross-section $S_h$~$=$~$\pi R_h^2$, and characteristic impedance $Z_{ch}= \rho_0 c_0 / S_h$.
The acoustic field in the side hole is described by the pressure at the bottom of the hole $p_{hb}$ and at the top of the hole $p_{ht}$.

A second tube of length $L_2=166$~mm and cross-section $S$ is branched downstream from the side hole.
The acoustic field in this tube is described by the pressure at the left extremity $p_2$ and by the pressure at the right extremity $p_{end}$.

\subsection{Viscothermal losses}
Viscothermal losses are introduced through the complex wavenumber $\Gamma_i(s)$, where $s$ is the Laplace variable and $i=\{1,2,h \}$ refers to the index of the tube considered. 
The function $G_i(s)$ is defined, such that
\begin{align*}
G_i(s) &= e^{-\Gamma_i(s) L_i} = \lambda_i e^{-\epsilon_i \sqrt{s}} e^{-\tau_i s},
\end{align*}
with
\begin{align*}
\lambda_i &= e^{-({\alpha_2 \ell_v L_i})/{R_i^2}}, &
\epsilon_i &= \frac{\alpha_1 L_i}{R_i}\sqrt{\frac{2 \ell_v}{c_0}}, & \tau_i &= \frac{L_i}{c_0}, 
\end{align*}
where $\alpha_1=1.044$,  $\alpha_2=1.080$,  and $\ell_v= 4 \cdot 10^{-8}~\mathrm{m}$ (Chap.\ 5.5 of \cite{bible2016}).

In practice, $G_i(s)$ are approximated by a first-order low-pass filter and a delay $\tilde{G}_i(s)$, following the work from \cite{guillemain2005real}.
Fractional delays are also accounted for through the order 1 filters proposed by \cite{laakso1996splitting}.

\subsection{Forward and backward-propagating pressure waves}\label{sec: model eqs}
In the following, time-domain variables are written in small letters ({e.g.}\ $p_2^+(t)$), and frequency-domain variables are written in capital letters ({e.g.}\ $P_2^+(s)$).

The propagation of the acoustic waves in the resonator is described through the forward and backward-propagating acoustic pressures $p^+$ and $p^-$.
They are related to the acoustic pressure and flow $(p, u)$ through the relationships:
\begin{align*}
p&= p^+ + p^-, & u &= \frac{p^+ - p^-}{Z},
\end{align*}
where $Z=Z_c$ in the main tube of cross-section $S$, and $Z= Z_{ch}$ in the side hole.

Since the tubes $L_1$, $L_2$ and $L_h$ are all cylindrical, the acoustic field can be described as  transmission lines equations in the frequency domain.
For the tube of length $L_1$:
\begin{align}
P_1^+ &= G_1 P_{in}^+, & P_{in}^- &= G_1 P_{1}^- .
\end{align}
For the tube of length $L_2$:
\begin{align}
P_{end}^+ &= G_2 P_{2}^+, & P_{2}^- &= G_2 P_{end}^- .
\end{align}
For the tube of length $L_h$:
\begin{align}
P_{ht}^+ &= G_h P_{hb}^+, & P_{hb}^- &= G_h P_{ht}^- .
\end{align}

\subsection{Boundary conditions}\label{sec:bcs}
The boundary conditions in the tube are described hereafter.

\subsubsection{Radiation}
First, radiation from the open end is neglected: the pressure $p_{end}$ is written consequently as
\begin{equation}
 p_{end} = 0.
\end{equation}

\subsubsection{Hole junction}
Secondly, since the register hole has a small diameter and a long chimney length, the series impedances of the hole can be neglected (section 3.3.2.2 of \cite{debut_analysis_2005}).
The boundary conditions at the bottom of the hole are therefore given by:
\begin{align}
p_1 &= p_2, \label{eq:p1p2}  \\ 
p_2 &= p_{hb}, \label{eq:p2ph}  \\ 
u_1 &= u_2 + u_{hb}. \label{eq:flowHole}
\end{align}

\subsubsection{Flow crossing the reed channel}
The next boundary condition involves $p_{in}$ and comes from the nonlinear characteristics of the flow entering the resonator. %\cite{dalmont2003nonlinear}.
In this relationship, the acoustic flow $u_{in}$ depends on the difference between the blowing pressure of the musician $p_m$ and the pressure at the input of the instrument $p_{in}$.
By assuming that the jet experiences total turbulent dissipation \cite{wilson_operating_1974} and modeling  the reed as a massless, undamped spring \cite{ollivier2005idealized}, the nonlinear characteristics is defined as \cite{dalmont2003nonlinear}:
\begin{equation}\label{eq:uin}
\hat{u}_{in} = \zeta [\hat{p}_{in} - \gamma +1]^+ \text{sgn}(\gamma - \hat{p}_{in}) \sqrt{|\gamma - \hat{p}_{in}|}, 
\end{equation}
where the function $[x]^+$ returns the positive-part of $x$, i.e.\ $ [x]^+ = (x + |x|)/2$.
The dimensionless blowing pressure is given by $\gamma= p_m/P_M$, where $P_M$ is the minimum pressure needed to close the reed channel in a quasi-static regime. 
Typical values of $P_M$ are in the range $P_M\in[4, 8]$~kPa, according to \cite{dalmont_oscillation_2007, atig2004saturation}.
The parameter $\zeta$ represents the embouchure, with common values for the clarinet between 0.05 and 0.4 \cite{dalmont_oscillation_2007}.
The dimensionless quantities are defined as
\begin{align*}
\hat{p}_{in}&=p_{in} / P_M, & \hat{u}_{in} &=  u_{in}Z_c/ P_M.
\end{align*}

In Eq.\ \eqref{eq:uin}, the dynamics of the reed are neglected to obtain a direct relationship between $p^+_{in}$ and $p^-_{in}$. 
This relationship is given in \cite{taillard2010iterated} and is detailed in the Appendix of \cite{bergeot2014response}.
It is expressed as:
\begin{align}\label{eq:Raman}
\hat{p}_{in}^+ = f_{\gamma \zeta}(\hat{p}_{in}^-)= \gamma - X[\gamma - 2 \hat{p}_{in}^-] - \hat{p}_{in}^-,
\end{align}
where the function $X$ is defined in Appendix A of \cite{taillard2010iterated}.

\subsubsection{Localized nonlinear losses in the register hole}
Localized nonlinear losses in the register hole are modeled using the following boundary condition for $p_{ht}$:
\begin{equation}\label{eq:bcnl pv}
p_{ht}(t) = \rho_0 C_\text{nl} v_{ht}(t)|v_{ht}(t)|,
\end{equation}
where $C_\text{nl}>0$ is the nonlinear losses coefficient, which depends on the roundness of the edges of the hole \cite{atig2004saturation}, and $v_{ht}$ is the acoustic speed at the top of the side hole. 
An explicit relationship between $p^+_{ht}$ and  $p^-_{ht}$ is given in \cite{szwarcberg2024oscillation}:
\begin{equation}
p_{ht}^-(t) = r_\mathrm{nl}\left[p_{ht}^+(t)\right],
\end{equation}
where
\begin{align}\label{eq:bcnl}
r_\mathrm{nl}(x) = x \left(1- \frac{4}{1+ \sqrt{1+K_\mathrm{nl}|x|}} \right),
\end{align}
with $K_\mathrm{nl}=8C_\mathrm{nl}/(\rho_0 c_0^2)$.
For $K_\mathrm{nl}=0$, we get $r_\mathrm{nl}(x)$~$=$~$-x$, which corresponds to an open hole boundary condition.
As $K_\mathrm{nl} \to \infty$, $r_\mathrm{nl}(x)=x$, meaning the hole is closed.

In a dimensionless form, $r_\mathrm{nl}$ is rewritten as $\hat{p}_{ht}^-$~$=$~$\hat{r}_\mathrm{nl}\left[\hat{p}_{ht}^+\right]$, where
\begin{align}\label{eq:bcnl adim}
\hat{r}_\mathrm{nl}(x) = x \left(1- \frac{4}{1+ \sqrt{1+\hat{K}_\mathrm{nl}|x|}} \right),
\end{align}
with $\hat{K}_\mathrm{nl}= P_M K_\mathrm{nl}=0.2$, assuming a moderate $P_M$ and a hole with sharp edges.

\subsection{Extraction of the modal acoustic pressure}
Modal acoustic pressures are useful to visualize the limit cycles of the different oscillating regimes.
However, they are not directly accessible through waveguide modeling. 
Filtering is applied \textit{a posteriori}, using  the modal decomposition of the input impedance $Z_{in}=P_{in} /U_{in}$:
\begin{equation}
Z_{in}   = Z_c\sum_n \frac{C_n}{s-s_n} + \frac{\mathrm{conj}(C_n)}{s-\mathrm{conj}(s_n)},
\end{equation}
where $C_n$ and $s_n$ are the complex residues and poles, computed through the residues theorem from the analytic definition of the input impedance.
In particular, the modal frequencies are given by $f_n=\Im(s_n)/(2\pi)$.
The $n$-th modal acoustic pressure at the input $p_n$ is defined through the following ODE:
\begin{equation}
\dot p_n(t) = Z_c C_n u_{in}(t) + s_n p_n(t),
\end{equation}\label{eq:modal}
where $\dot p_n=\partial_t p_n$ and $u_{in}=(p_{in}^+  - p_{in}^-)/Z_c$.
Modal acoustic pressures can then be computed by filtering $u_{in}$ with an IIR filter.

\section{Simulations}
Blowing pressure ramps are first carried out to find the ranges for which the first register is stable when the hole is closed, as well as the range for which the second register is stable when the hole is open.  

Multiple hole openings are then performed for constant control parameters to assess the playability of the  register jumps.

\subsection{Blowing pressure ramps}\label{sec:ramps}
The embouchure parameter is fixed at $\zeta=0.3$ (average value for clarinet playing \cite{dalmont_oscillation_2007}) throughout the simulations.
First, two blowing pressure ramps (\textit{crescendi}) are performed: one for the hole closed ($\hat K_\mathrm{nl} \to \infty$), one for the hole open ($\hat K_\mathrm{nl}=0.2$).
For both cases, the values of the blowing pressure for which oscillations start ($\gamma_\mathrm{osc}^{(c)}, \gamma_\mathrm{osc}^{(o)}$) and stop ($\gamma_\mathrm{ext}^{(c)}, \gamma_\mathrm{ext}^{(o)}$) are noted.
Multistability zones are then determined, knowing as an  inner-property of reed instruments, that the equilibrium (no sound, noted R0) is stable when $\gamma>1$ \cite{dalmont2005analytical}.

\subsection{Hole openings}\label{sec:holeopenings}
From a constant blowing pressure $\gamma$, the hole is instantaneously opened.
The frequency of $p_{in}$ is computed before the opening of the hole ($f^{(c)}$), and after ($f^{(o)}$).
The ratio $f^{(o)}/f^{(c)}$ determines the register obtained. 
In particular, if $f^{(o)}/f^{(c)}\approx 3$, the second register is played.

For one value of the blowing pressure, $N_o=200$ hole openings are performed at different times distributed over a time window $T^{(c)}=1.5/f_1^{(c)}$, where $f_1^{(c)}$ is the first modal frequency for the closed hole.
The value of $T^{(c)}$ is chosen to ensure that all limit cycles can be fully sampled.
The proportion of the second register obtained for the $N_o$ openings is computed.

Four thresholds are measured through this procedure: $\gamma_\mathrm{min}^{(0\%)}, \gamma_\mathrm{min}^{(100\%)}, \gamma_\mathrm{max}^{(100\%)}, \gamma_\mathrm{max}^{(0\%)}$.
Thresholds $\gamma_\mathrm{min}^{(100\%)}$ and $ \gamma_\mathrm{max}^{(100\%)}$ denote the minimum and maximum blowing pressures that always lead to the second register (R2) when the hole is opened.
$\gamma_\mathrm{min}^{(0\%)}$ is the maximum blowing pressure lower than $\gamma_\mathrm{min}^{(100\%)}$  that never leads to R2, and $\gamma_\mathrm{max}^{(0\%)}$ is the minimum blowing pressure greater than $\gamma_\mathrm{max}^{(100\%)}$ that never leads to R2.

To explore if the initial conditions that lead to a specific regime are sensitive to a small perturbation, a random impulsion is added to $\hat p_{in}^+$ when the hole is opened.
Different amplitudes of perturbation are tested, between $0.0$ and $0.50$.

\section{Results}

\subsection{Blowing pressure ramps}
\begin{figure}
	\centering
	\includegraphics[width=.5\textwidth]{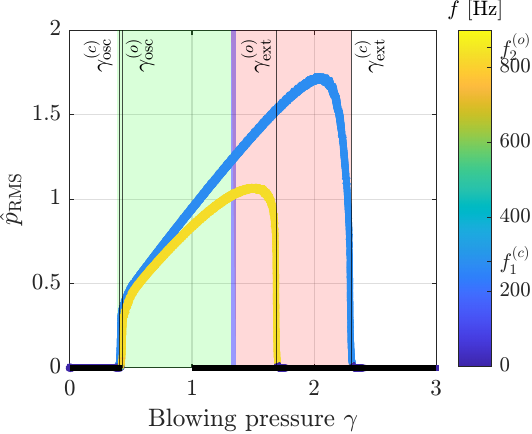}
	\caption{Evolution of the amplitude of the acoustic pressure into the mouthpiece when the blowing pressure $\gamma$ increases linearly , for a constant embouchure $\zeta=0.3$. 
	Two \textit{crescendi} are represented: one for the hole closed (blue curve), and one for the hole open (yellow curve).
	Colored surfaces in the background show the ranges of $\gamma$ where the second register is reached with a given probability when the hole is opened.
	Green: $100~\%$.
	Blue: between $0~\%$ and $100~\%$.
	Red: $0~\%$.
	}
	\label{fig:rampsim}
\end{figure}

Figure \ref{fig:rampsim} shows the evolution of the amplitude of the acoustic pressure in the mouthpiece $\hat p_\mathrm{RMS}$, as the blowing pressure increases linearly from $\gamma=0$ to $\gamma=3$ over 20~s.
The blue and yellow curves represent the cases with the hole closed and open, respectively. 
The color mapping corresponds to the oscillation frequency: for the closed hole, the playing frequency is near the first modal frequency $f_1^{(c)}$ (first register R1), while for the open hole, it is close to the second modal frequency $f_2^{(o)}$ (second register R2).

For each case, the minimum $(\gamma_\mathrm{osc}^{(c)}, \gamma_\mathrm{osc}^{(o)})$ and maximum blowing pressures $(\gamma_\mathrm{ext}^{(c)}, \gamma_\mathrm{ext}^{(o)})$ that sustain oscillations are measured.
The thick black lines at the bottom of the graph indicate values of $\gamma$ where the equilibrium (no sound, R0) is stable.
Note that the values of $(\gamma_\mathrm{osc}^{(c)}$ and $\gamma_\mathrm{osc}^{(o)})$ are overestimated due to bifurcation delay \cite{bergeot2013prediction}.

For the open hole, when $\gamma \in [1, \gamma_\mathrm{ext}^{(o)}]$ R0 and R2 are both stable.
In this range, if a musician plays in the first register with the hole closed and then opens the register hole, they may end up in either R2 or R0.
The hole opening procedure described in Section~\ref{sec:holeopenings} is used to evaluate how  the probability of playing R2 evolves with $\gamma$.
The four threshold values characterizing this probability are listed in Table~\ref{tab:thresh}.

\begin{table}
\centering
\caption{Values of the four thresholds characterizing the probability of playing the second register when opening the hole.}
\label{tab:thresh}
\begin{tabular}{lcccc}
\hline
Threshold & $\gamma_\mathrm{min}^{(0\%)}$ & $\gamma_\mathrm{min}^{(100\%)}$ & $\gamma_\mathrm{max}^{(100\%)} $& $\gamma_\mathrm{max}^{(0\%)}$ \\
Value & $0.383$ & $0.386$ & $1.32$ & $1.36$ \\
\hline
\end{tabular}
\end{table}

Within the green region, R2 is always reached. 
In the blue region around $\gamma=1.3$, the probability of playing R2 depends on the timing of the register hole opening. 
In the red region, the system always lands on the R0.
Hence, the range of $\gamma$ in which R2 can be played after opening the register hole from R1 is narrower than the range of $\gamma$ where R2 is stable.

To characterize the transition regions, $\gamma \in [\gamma_\mathrm{min}^{(0\%)}, \gamma_\mathrm{min}^{(100\%)}]$ and $\gamma \in [\gamma_\mathrm{max}^{(100\%)}, \gamma_\mathrm{max}^{(0\%)}]$, multiple hole openings at different times are needed to determine which regime the model will predominantly converge to. 
Compared to the green or red regions, characterizing the blue region requires $N_o$ times more simulations.
For instrument makers studying the playability of twelfths in a clarinet model, the narrowness of these transition regions is an encouraging result.

%pour le guide d'onde étudié, c'est une bonne nouvelle que la zone bleue soit si petite : dans cette zone de multistabilité, si l'on effectue des ouvertures de trou toujours au même moment, on pourrait passer à côté du caractère multistable du système. en particulier, on pourrait penser que le deuxième registre n'est pas jouable, alors qu'en ouvrant le trou à un autre moment on parvienne à jouer du deuxième registre. 

A closer look at the transition from $\gamma_\mathrm{max}^{(100\%)}$ to $\gamma_\mathrm{max}^{(0\%)}$ is carried out in the next section.

\begin{figure}
	\centering
	\includegraphics[width=.5\textwidth]{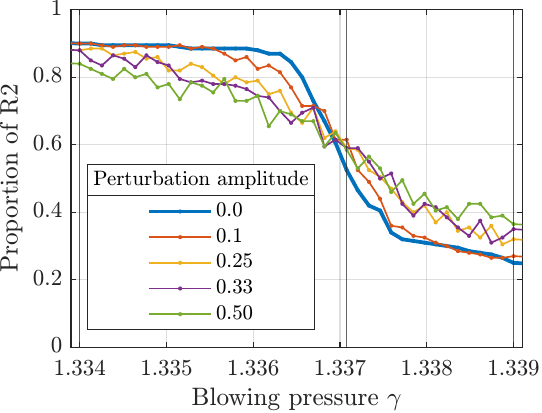}
	\caption{Evolution of the proportion of the second register (R2) when opening the register hole, for $\gamma\in[1.334,1.339]$. 
	The red, yellow, purple and green curves show the evolution of the proportion of R2 when random perturbations are added to $\hat p^+_{in}$ when the register hole is opened.}
	\label{fig:propR2}
\end{figure}

\subsection{Evolution of the probability of playing the second register}

Figure \ref{fig:propR2} shows how the probability of playing the second register evolves within $\gamma\in [\gamma_\mathrm{max}^{(100\%)} , \gamma_\mathrm{max}^{(0\%)}]$.
The probability follows negative sigmoid curve: it decreases gradually from $100~\%$ to $90~\%$, then drops sharply between $\gamma=1.336$ and $\gamma=1.338$.

\begin{figure*}
	\centering
	\includegraphics[width=\textwidth]{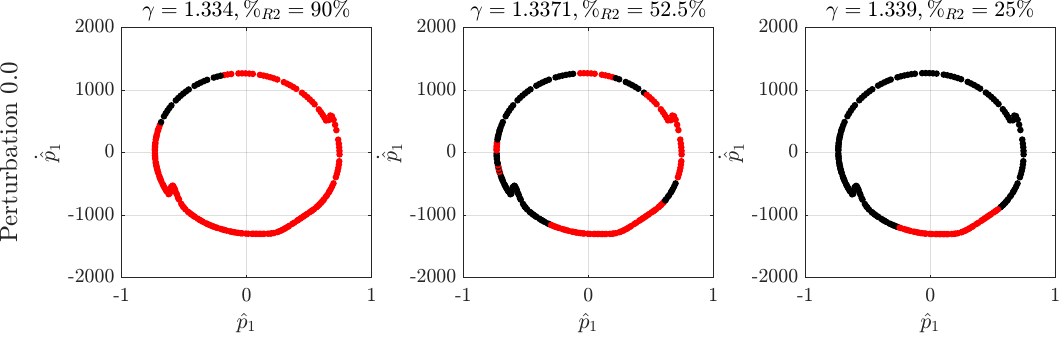}
	\includegraphics[width=\textwidth]{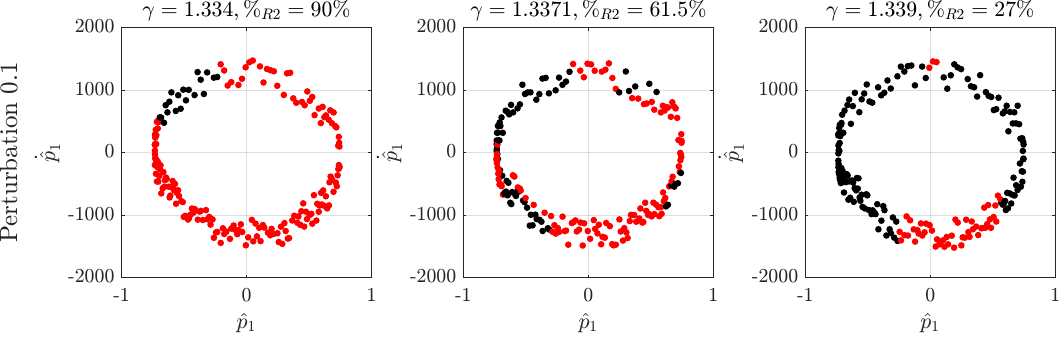}
	\includegraphics[width=\textwidth]{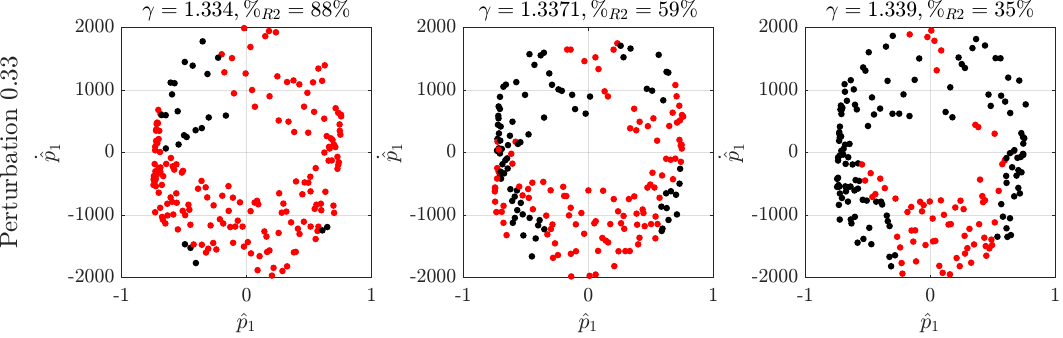}
	\caption{Positions on the limit cycle of the first register leading, when the hole is opened, to the second register (in red), and to the equilibrium (in black).
	Limit cycles are represented in the $(\hat p_1, \dot{\hat p }_1)$ space.
	Three different values of $\gamma$ are displayed on each row.
	On the second row, a random perturbation of $\pm 0.1$ is added to $\hat p_{in}^+$ when the hole is opened; on the third row the amplitude is $0.33$.	
	}
	\label{fig:lc}
\end{figure*}

In the projection of the phase space on the variables $(\hat p_1, \dot{ \hat p }_1)$, Figure \ref{fig:lc} illustrates how the system converges to either R0 or R2 depending on the position of the initial condition on the R1 limit cycle.
For instance, for $\gamma=1.334$ (top left panel), only initial conditions within a small section of the cycle (around an angle of $3\pi/4$) are attracted to R0 (black dots).
Opening the hole at a different time, corresponding to another position on the limit cycle, leads to R2 (red dots).
This behavior is an instance of phase-tipping \cite{alkhayuon2021phase}, where the result of a transition depends not only on the perturbation but also on the phase at which it is applied.
As $\gamma$ increases and the probability of playing R2 decreases, the set of initial conditions leading to R0 expands. 

The top-center panel of Figure \ref{fig:lc} shows that the set of initial conditions leading to R0 is not necessarily connected, highlighting the complex structure of the basin of attraction of the equilibrium in the $(\hat p_1, \dot{ \hat p }_1)$ plane.
At $\gamma=1.339$ (top right panel), only a small portion of the limit cycle, around an angle of $-\pi/3$, leads to R2.

This visualization suggests that opening the register hole at the same time in every simulation is not ideal.
Within $\gamma \in [\gamma_\mathrm{max}^{(100\%)} , \gamma_\mathrm{max}^{(0\%)}]$, if the hole were always opened when $(\hat p_1, \dot{ \hat p }_1) \approx (0,-1200)$, the model would consistently produce R2.
However, at $\gamma=1.339$ (top right panel) there are few occurrences of R2.

Introducing a small random perturbation to $\hat p^+_{in}$ when opening the register hole results in smoother sigmoid curves in Figure \ref{fig:propR2}, indicating that the basins of attraction of R0 and R2 remain stable under perturbations below 0.1. 
The second row of Figure \ref{fig:lc} supports this, revealing additional regions belonging to R2’s basin, such as the red dots around $\pi/2$ for $\gamma=1.339$.
Larger perturbations blur the basin boundaries in the $(\hat p_1, \dot{ \hat{p} }_1)$ plane, as seen in the third row of Figure \ref{fig:lc}.

To better assess the basin stability of R0 and R2 when opening the register hole, alternative methods could be explored. 
Since perturbing $p_{in}^+$ primarily affects $\dot{\hat p}_1$, adding white noise to the blowing pressure $\gamma$ and the embouchure parameter $\zeta$ could provide a more balanced perturbation across the phase space.

\section{Conclusion}
A waveguide model of a clarinet with a register hole is studied. 
The blowing pressure ranges for which the first and the second registers are stable are determined. 
In particular, for the open hole configuration, the multistability regions of the second register and the equilibrium are quantified.

Results indicate that within this multistability range, repeatedly opening the register hole from the first register can lead the system to either the equilibrium or the second register. The probability of reaching a given register follows a sigmoid-shaped evolution as the blowing pressure increases. This behavior is reflected in phase space by the structure of the basins of attraction, which progressively enclose the limit cycle of the first register. The shifting probability of convergence to a given regime directly corresponds to changes in the shape of these basins.

Finally, the robustness of the basins of attraction is assessed by introducing random perturbations to the initial conditions. Preliminary results suggest that the basins remain robust under perturbations of $p_{in}^+$ smaller than 0.1.

\section{Acknowledgments}
This study has been supported by the French ANR LabCom LIAMFI (ANR-16-LCV2-007-01). 

\bibliography{biblio}

\end{document}